\documentclass[sigconf]{acmart}

\usepackage{tabularx}
\usepackage{multirow}
\usepackage{enumitem}
\usepackage{url}
\usepackage{wrapfig}
\usepackage{booktabs}

\copyrightyear{2020} 
\acmYear{2020} 
\setcopyright{rightsretained} 
\acmConference[SAC '20]{The 35th ACM/SIGAPP Symposium on Applied Computing}{March 30-April 3, 2020}{Brno, Czech Republic}
\acmBooktitle{The 35th ACM/SIGAPP Symposium on Applied Computing (SAC '20), March 30-April 3, 2020, Brno, Czech Republic}
\acmDOI{10.1145/3341105.3374114}
\acmISBN{978-1-4503-6866-7/20/03}
\acmArticle{4}
\acmPrice{15.00}

\AtBeginDocument{%
  \providecommand\BibTeX{{%
    \normalfont B\kern-0.5em{\scshape i\kern-0.25em b}\kern-0.8em\TeX}}}

\begin{document}
%% =========================================================================================

\title[Unsupervised Cross-Modal Audio Representation Learning from Audio and Text]{Unsupervised Cross-Modal Audio Representation Learning from Unstructured Multilingual Text}

%% =========================================================================================

\author{Alexander Schindler}
\affiliation{%
  \institution{Center for Digital Safety and Security \\ Austrian Institute of Technology}
  \streetaddress{Giefinggasse 4}
  \city{Vienna}
  \country{Austria}
  \postcode{1210}
}
\email{alexander.schindler@ait.ac.at}

\author{Sergiu Gordea}
\affiliation{%
  \institution{Center for Digital Safety and Security \\ Austrian Institute of Technology}
  \streetaddress{Giefinggasse 4}
  \city{Vienna}
  \country{Austria}
  \postcode{1210}
}
\email{sergiu.gordea@ait.ac.at}

\author{Peter Knees}
\affiliation{%
  \institution{Faculty of Informatics \\ TU Wien}
  \streetaddress{Favoritenstraße 9-11/194-1}
  \city{Vienna}
  \country{Austria}
  \postcode{1040}
}
\email{peter.knees@tuwien.ac.at}

%% =========================================================================================

\renewcommand{\shortauthors}{Schindler, Gordea and Knees}

%% =========================================================================================
\begin{abstract} 

We present an approach to unsupervised audio representation learning. Based on a triplet neural network architecture, we harnesses semantically related cross-modal information to estimate audio track-relatedness. By applying Latent Semantic Indexing (LSI) we embed corresponding textual information into a latent vector space from which we derive track relatedness for online triplet selection. This LSI topic modelling facilitates fine-grained selection of similar and dissimilar audio-track pairs to learn the audio representation using a Convolution Recurrent Neural Network (CRNN). By this we directly project the semantic context of the unstructured text modality onto the learned representation space of the audio modality without deriving structured ground-truth annotations from it. We evaluate our approach on the Europeana Sounds collection and show how to improve search in digital audio libraries by harnessing the multilingual meta-data provided by numerous European digital libraries. We show that our approach is invariant to the variety of annotation styles as well as to the different languages of this collection. The learned representations perform comparable to the baseline of handcrafted features, respectively exceeding this baseline in similarity retrieval precision at higher cut-offs with only 15\% of the baseline's feature vector length.
 
\end{abstract}
%% =========================================================================================




%% =========================================================================================

\maketitle

% =================================================================
\section{Introduction}
\label{intro}
% =================================================================

Audio representations aim to capture intrinsic properties and characteristics of the audio content to facilitate complex tasks such as classification (acoustic scenes \cite{giannoulis2013detection,Lidy2016}, music genres \cite{lidy2007improving}), regression (emotion recognition \cite{yang2012machine}) or similarity estimation (music,\cite{knees2016music} general audio \cite{schindler2019large}). In the context of this paper we focus on their application in audio similarity estimation and retrieval. More specifically, for heterogeneous audio collections provided by digital libraries in the cultural heritage domain.

\begin{figure}[t]
  \centering \includegraphics[width=0.5\textwidth]{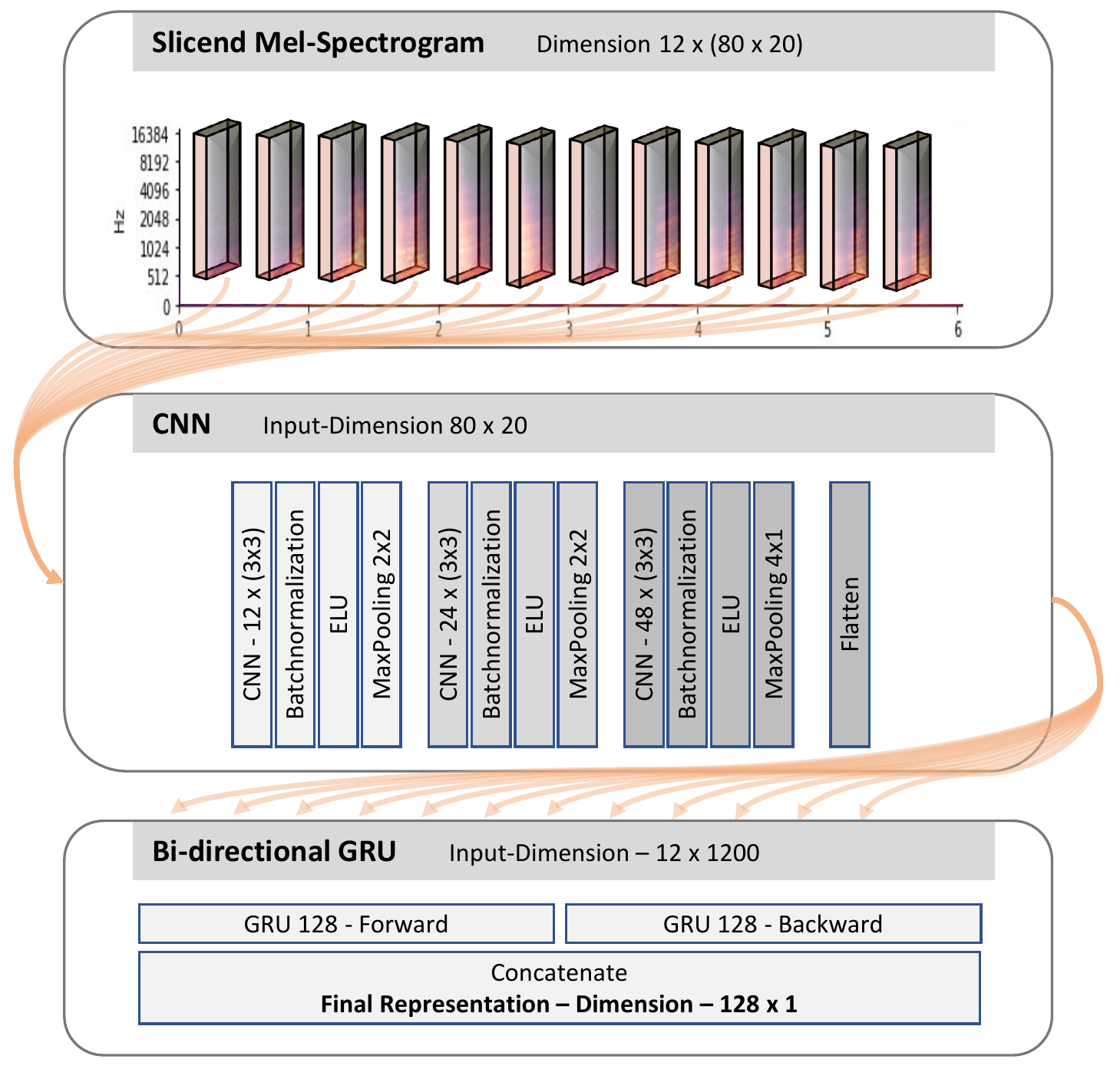}
  \caption{Model Overview: Convolutional Recurrent Neural Network (CRNN). a) the input Mel-Spectrogram (80x130) is split into 12 segments (80x12, 50\% overlap). b) each segment is processed by a shared CNN stack resulting in a sequence of 12 intermediate embedding vectors. c) the sequence is processed by a bi-directional GRU and its output is the learned representation (256x1).}
  \label{fig:modelarchitecture}
\end{figure}

% digital libraries are special
Digital libraries (DL) present unique challenges for information retrieval research. The information need of DL users is highly specific and users are often highly experienced within the search domain. The challenging requirement for effective tools to search and discover content in large databases faces major obstacles such as heterogeneity, multi-modality and often multi-linguality of the content stored in these databases. The common approach taken to satisfy these information needs is to provide as much, rich and accurate meta-data as possible and to apply information retrieval and semantic computing \cite{raimond2007music} technologies to index this meta-data to facilitate efficient search. Either of these approaches requires considerable amounts of manual interaction to annotate or interlink data items and does not scale well.
Approaches based on meta-data require that DL users are acquainted with the correct terminology used to describe the item or the categories. Based on the fact that many collections in DLs are aggregated and curated from scientists from specific research disciplines such as history, archaeology or musicology, this terminology can be very specific and not everyone might be familiar with it on the same level. Further, not every type of information can be efficiently described using textual meta-data. Content related relations such as ``sounds like'' are highly complex and difficult to describe by meta-data.
This heterogeneity is also a challenge for the definition and modeling of the \textit{acoustic similarity} function. The Europeana\footnote{\url{https://www.europeana.eu}} Sounds data-set \cite{schindler2016europeana} contains besides \textit{Music} also \textit{Spoken Word} in form of interviews, radio news broadcast, public speeches, field recorded \textit{Animal-} and \textit{Ambient-} or \textit{Environmental-Sounds}. Additionally, the recordings vary in instrumentation and recording quality (from digitized wax-tapes to born-digital content).
In \cite{schindler2016europeana} we applied a diverse set of handcrafted audio- and music-descriptors to model an audio-content based similarity estimation function for the \textit{Europeana} data. The most critical part of this approach was the selection of the features to adequately describe the heterogeneous semantics of the different collections in the data-set, as well as the balancing of the feature weights to approximate the subjective similarity estimation. Feature weight optimization was approached empirically through a predefined set of similar records. During an iterative process the weights of the different features were adapted.
This manual optimization process is sub-optimal in terms that it only optimizes towards a small set of manually selected items. An optimization against the entire data-set would require that pairwise similarity estimations would be available as ground-truth-assignments. 
Because creating such assignments is not feasible on large scales, a common approach is to define similarity by categorical membership or identity. In \cite{park2018representation} the authors defined music similarity on tracks originating from the same artists and used a triplet deep neural network architecture (see Section \ref{fig:modelarchitecture}) to learn an optimized music optimization. Using identity or categorical data as ground-truth is sub-optimal \cite{schindler_knees:cbmi:2019} because it usually defines acoustic similarity too coarse.
In \cite{schindler_knees:cbmi:2019} we learned a music representation from multi-label assignments using Latent Semantic indexing (LSI) to project the discrete information into a continuous space from which a track-similarity function based on tag-relatedness was derived. This tag-based track-similarity was then transferred to the audio space using a Triplet Network with a margin maximizing loss and online triplet selection strategy. 
Following this approach a network learns to maximize the margin between the distances of a reference track and its positive similar example and its negative dissimilar example, in a learned semantic embedding space, based on the constraint, that the vector distance of the positive pair should be much smaller than the distance of the negative pair.
The challenge is the selection of positive and negative examples to a reference track. 
Still, these approaches rely on ground-truth assignments for supervised learning of audio representations. 

In this paper we build on the conclusions of \cite{schindler_knees:cbmi:2019} that by adding more content related information to the input space of the LSI projections, the definition of LSI topics improves, resulting in a better track-relatedness function. We adopt this approach and extend it from a discrete categorical space with a fixed vocabulary to an open, unstructured, multi-lingual free-text space. The meta-data provided by professional librarians contains such item-related descriptions including descriptions of audible content. The major contribution of this approach is, that it uses unstructured text to derive track-relatedness and does not require structured ground-truth assignments.
To demonstrate our approach we first discuss and position our approach within related work in Section \ref{sec:relatedwork} before we describe our method in Section \ref{sec:method}. We extensively evaluate the approach in Section \ref{sec:evaluation} and thoughtfully discuss the results in Section \ref{sec:results} before we draw conclusion and discuss future work in Section \ref{sec:conclusions}.

% =================================================================
\section{Related Work}
\label{sec:relatedwork}
% =================================================================

\subsection{Content Retrieval in Music Digital Libraries} 

Concerning Music Digital Libraries (DL) this paper is mostly related to \cite{schindler2016europeana} where we presented an approach to audio-content similarity estimation within Europeana Sounds project. Following a late-fusion approach, the weighted combination of different audio-content descriptors was applied to highly heterogeneous data. The presented evaluation method is also adopted in this paper. Issues of data aggregation in audio DLs are described in \cite{koch2009}.

\subsection{Music Similarity Retrieval} %OK

Search-by-example, such as finding music tracks that are similar to a query track, is an actively researched task \cite{knees2015music,knees2016music}. Research on music similarity estimation currently faces two major obstacles. First, music similarity is a highly subjective concept and is strongly influenced by the listening habits and music taste of the listener \cite{berenzweig2004large}.
Second, state-of-the-art approaches to music similarity estimation are still not able to satisfactorily close the semantic gap between the computational description of music content and the perceived music similarity \cite{knees2013survey}.
The many facets of music similarity - such as concrete music characteristics (e.g. rhythm, tempo, key, melody, instrumentation), perceived mood (e.g. calm, aggressive, happy), listening situation (e.g. for dinner, to concentrate, for work out), musicological factors (e.g. composer influenced by) - complicate the definition of a unified music description which captures all semantic music concepts.
Traditionally this has been approached by defining a set of features, which extract certain low level music characteristics such as timbre \cite{logan2001music} or rhythm \cite{lidy2005evaluation}, mid-level properties such as chords \cite{muller2015fundamentals}, but also high-level features. This approach faces the problem that hand-crafted feature design is neither scaleable nor sustainable \cite{humphrey2013feature}. 
Representation learning using Deep Neural Networks (DNN) has been actively explored in recent years \cite{sigtia2014improved,schroff2015facenet} as an alternative to feature engineering.
Although some of these approaches outperform feature-based methods, a major obstacle is their dependency on large amounts of training data.
Although it has been shown that shallow DNNs have an advantage on small datasets \cite{Schindler2016} they struggle to describe the latent complexity of music concepts and do not generalize on large datasets \cite{humphrey2013feature}. 

% === Representation Learning ===
\subsection{Representation Learning (RL)} %OK

RL using DNNs gained attention through the publication of \textit{FaceNet} \cite{schroff2015facenet} which significantly improved the state-of-the-art of face re-identification.
This approach is based on global item relatedness where faces are similar when they belong to the same person and otherwise are dissimilar.
A similar approach using the global relatedness of performing artists has been applied to music data \cite{park2018representation}.
Contextualized relatedness especially in the domain of music has been used in \cite{Sandouk2016}.
A similar approach to this paper of estimating tag-relatedness from user-tags was taken in \cite{Sandouk2016}. Latent Dirichlet Analysis (LDA) was used to project the categorical data into a numerical space. The approach was evaluated using a siamese neural network on three smaller datasets including the MSD subset using the noisy user-generated tag-sets of the Last.fm dataset. A differentiated evaluation of the learned semantic context as provided in this paper was missing.
%
% we should reference the CBMI paper somehow at least...
Concerning Representation Learning the presented paper in mostly related to \cite{schindler_knees:cbmi:2019} in which we extended the Million Song Dataset (MSD) \cite{Bertin-Mahieux2011} with additional ground truth multi-label assignments for \textit{Moods}, \textit{Styles} and \textit{Themes}. Further, we extended the single-label \textit{Genre} labels provided in \cite{schindler2012} to multi-label assignments. In \cite{schindler_knees:cbmi:2019} a music representation was learned from these multi-label assignments using Latent Semantic indexing (LSI) to project the categorical information into a continuous space from which a track-similarity function based on tag-relatedness was derived. This tag-based track-similarity was then transferred to the audio space using a Triplet Network with a margin maximizing loss and online triplet selection strategy.

% =================================================================
\section{Method}
\label{sec:method}
% =================================================================

% ----------------------------

The proposed method is based on a triplet neural network architecture to learn the contextualized semantic representation using a max-margin hinge loss with online triplet selection.

\subsection{Representation Learning}

To learn the acoustic representation we use a triplet network based on a shared Convolutional Recurrent Neural Network (CRNN) architecture \cite{choi2017convolutional}. The base-model is described in Section \ref{modelarchitecture} and depicted in Figure \ref{fig:modelarchitecture}. Using this triplet network, an input audio spectrogram $x$ is embedded $f(x_{i}^{a})$ into a  $d$-dimensional Euclidean space $\mathbb{R}^{d}$. The input consists of a triplet of audio content items: a query track (anchor) $x_{i}^{a}$, a track similar (positive) $x_{i}^{p}$ and one dissimilar (negative) $x_{i}^{n}$ to the query. The objective is to satisfy the following constraint:

\begin{equation} \label{eq1}
   \left \| f(x_{i}^{a})-  f(x_{i}^{p}) \right \|_{2}^{2} + \alpha < \left \| f(x_{i}^{a}) - f(x_{i}^{n}) \right \|_{2}^{2}  
\end{equation}

\noindent
For $\forall ( f(x_{i}^{a}), f(x_{i}^{p}), f(x_{i}^{n}) ) \in \tau $, where $\left \| f(x_{i}^{a})-  f(x_{i}^{p}) \right \|_{2}^{2}$ is the squared Euclidean distance between $x_{i}^{a}$ and $x_{i}^{p}$, which should be much smaller than the distance between $x_{i}^{a}$ and $x_{i}^{n}$. $\alpha$ is the enforced margin between positive and negative pair-distances. $\tau$ represents the set of all possible triplets in the training-set. The objective of Eq. \ref{eq1} is reformulated as the following triplet-loss function:

\begin{equation} \label{eq3}
\sum_{i=1}^{N} \max \left [ \left \| f(x_{i}^{a}) - f(x_{i}^{p}) \right \|_{2}^{2} - \left \| f(x_{i}^{a}) - f(x_{i}^{n}) \right \|_{2}^{2} + \alpha \right ]
\end{equation}

\subsubsection{Online Triplet Selection} 

Efficient selection of triplets is a crucial step in training the network. Generating all possible triplet combinations $\tau$ is inefficient due to the cubic relation and the lacking contribution to the training-success of triplets not violating Eq. \ref{eq1}. Thus it is required to select hard triplets violating this constraint.
A common approach to this is \textit{online triplet selection} where triplets are combined on a mini-batch basis \cite{schroff2015facenet}.
To select appropriate triplets the batch size needs to be appropriately large. We use a batch size of 400 tracks. Their corresponding log-scaled Mel-Spectrograms (see Sec. \ref{dataprep}) are embedded into a latent space. 
%
% LSI => positive/negative pairs
The selection of positive and negative examples is based on the semantically embedded textual information extracted from the meta-data (explained in more detail in Sec. \ref{sec:method:tagsim}). The pairwise cosine-distance $\cos(LSI_{1}^{ts}, LSI_{2}^{ts})$ matrix of the corresponding $l_{2}$ normalized LSI-embeddings is calculated for all mini-batch instances. The diagonal elements are set to zero to avoid identical pairs. Thresholds for pair-selection were evaluated empirically by analyzing the distribution of the cosine-distance space of the LSI-embeddings and set to $\cos(LSI_{1}^{ts}, LSI_{2}^{ts}) \geq 0.8$ (upper) and $\cos(LSI_{1}^{ts}, LSI_{2}^{ts}) < 0.5$ (lower).
For each row in the LSI-embeddings similarity matrix that contains valid positive and negative instances, the squared Euclidean distances of the corresponding audio embeddings are calculated.
$argmin$ is computed to identify relevant positive and negative pairs. This deviates from the original approach \cite{schroff2015facenet} where $argmin$ is used to select hard negatives and $argmax$ for hard positive pair examples. The intention of the approach presented in \cite{schroff2015facenet} is to be be invariant to image background as well as to changes in pose, color and illumination. This is supported by their hard triplet selection method which enforces to learn highly discriminate object-related features. The ``sounds like'' audio similarity of in this paper defines a global similarity also taking ``background noise'' into account. This is emphasized by using $argmin$ to select tracks which similar features in the embedding space as positive pairs. Finally, instances where no positive and negative example are found are removed.

\begin{figure*}[t]
  \centering \includegraphics[width=1.0\textwidth]{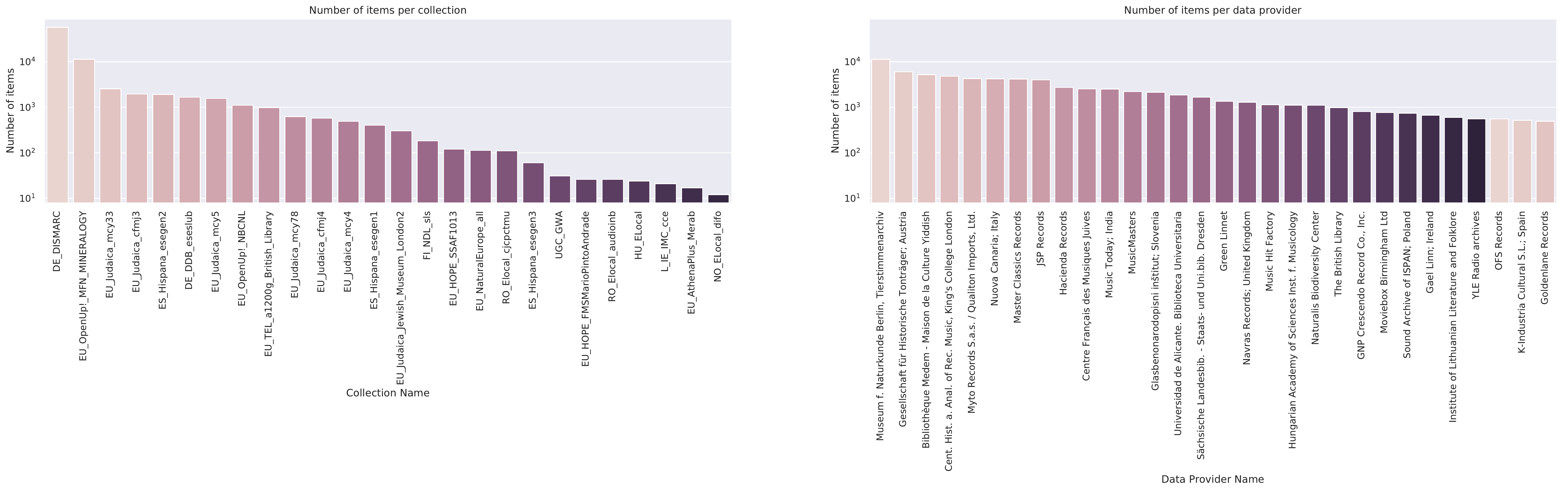}
  \caption{Data-set overview: left) Number of items per collection. right) number of items per data provider. Y-axis is log-scaled for both charts due to unbalanced frequency distributions.}
  \label{fig:dataoverview}
\end{figure*}

\subsection{Relatedness Measure}
\label{sec:method:tagsim}
% ----------------------------
% LSI
% ----------------------------
%goal ->audio-content based similarity estimation

To train the triplet network with target values reflecting the similarity of two records, we build a measure capturing the similarity of their associated meta-data descriptions.
To this end, we make use of all the meta-data entries in all categories (see Fig.~\ref{fig:datadensity}) and apply text processing to their concatenation, i.e. a standard word-level \emph{term frequency by inverse document frequency} (TFIDF) scheme emphasizing terms specific to individual record meta-data in favor of meta-data entries common to many records. % add specific formula here!?
This should lower the impact of collection specific keywords on the learned audio representation while not fully discarding this underlying relationship. 
To combat the facts that meta-data entries are often correlated and/or applied inconsistently across different collections, i.e., effects similar to \emph{synonymy} and \emph{polysemy} in natural language processing, respectively, as a next step, we perform Latent Semantic Indexing (LSI)~\cite{deerwester:lsa:1990}.
LSI models latent topics as semantic associations of terms in a corpus of documents (in our case the individual words over all metadata of all records weighted according to the TFIDF scheme).
%From the given corpus, clusters of co-occurring terms belonging to a latent topic can be uncovered.
Technically, LSI operates on the $m\times n$ weight matrix $W$ with $w_{ij} = tfidf_{ij}$, where each row corresponds to a term $t_{j=1\ldots m}$ and each column to a record $a_{i=1\ldots n}$.
Latent topics---and with it an implicit spectral clustering---are derived by performing truncated singular value decomposition (SVD) for approximation of $W$.
As a result, each individual record is represented in the LSI-derived concept space via a cluster affinity vector (in the following referred to as LSI vector), cf.~\cite{Knees2016}.
A characteristic of SVD is that emerging topics are sorted in order of decreasing importance wrt. reconstruction of the data.
Therefore, the number of considered topics $l$ (referring to the first $l$ dimensions of the LSI vectors) can be used to steer the trade-off between generalization of the model and preservation of the original meta-data information in $W$. %, with low values resulting in a coarse approximation and higher numbers  
%We experiment with values of up to $400$ dimensions.

For calculation of record relatedness using the meta-data, we calculate the cosine similarity between the records' LSI vectors. 
From this, we sample positive and negative examples to be presented to the triplet network. 

\section{Evaluation}
\label{sec:evaluation}
% =================================================================

% OK
The aim of this evaluation is to asses if our method facilitates to harness semantic information from the meta-data space to learn a corresponding, general applicable audio representation from it. To show this, we perform three experiments using the task of audio similarity retrieval with the following settings:

\begin{enumerate}
    \item \textbf{Baseline:} selection of weighted handcrafted features \cite{Schindler2016} intended to show how neural network based approaches compare against state-of-the-art handcrafted feature-sets reported in literature.
    
    \item \textbf{Track-relatedness by collection:} triplet-based neural network using collection membership for online triplet selection. Acts as a second baseline, representing audio representation learning approaches relying on categorical data for triplet-selection. We therefor use collection membership to select positive and negative track-pairs.
    
    \item \textbf{Track-relatedness by LSI similarity:} our approach - triplet-based neural network using LSI-vector similarity for online triplet selection.
    
\end{enumerate}

We perform controlled experiments. The same model architecture as described in the following subsection is used for all experiments. We further take control over all random processes such as kernel initialization, shuffling of training instances after each epoch to reduce random effects and variance of the experimental results. The same training, validation and test splits are used in all experiments. By controlling all these parameters to our best knowledge we hypothesize that the learned representations are only influenced by LSI representation of the semantic space of the textual meta-data.

% =================================================================
\subsection{Model Architecture} % OK
\label{modelarchitecture}

For the evaluation we are using a Convolutional Recurrent Neural Network (CRNN) \cite{choi2017convolutional,schindler2019large}. A CRNN is a combination of a Convolutional Neural Network (CNN) stack and a Recurrent Neural Network (RNN). The CNN learns to identify patterns in the local 2D space of the input Spectrograms. The resulting feature transformation is passed on to the RNN which identifies sequential patterns in this intermediate embedding space. In our model the context learned by the RNN represents the final learned audio representation.

The model concept and architecture is depicted in Figure \ref{fig:modelarchitecture}. Instead of globally pre-normalizing the input-space, we use a BatchNormalization layer on top of our model.This normalized input matrix of shape 80x130 is then split into 12 sequential segments of shape 80x20 which overlap by 50\% (see top of Fig. \ref{fig:modelarchitecture}). Each segment is then processed by the CNN stack which consists of three blocks - each one containing a convolution layer with $3x3$ filter units, BatchNormalization followed by an Exponential Linear Unit (ELU) activation function and MaxPooling to down-sample the feature-maps. For the specific parametrization of the layers please refer to Fig. \ref{fig:modelarchitecture}). The feature-maps of the last block are flattened to a feature vector representing the intermediate non-linear feature transformation. The concatenation of all 12 feature vectors serves as input to a bi-directional Gated Recurrent Unit (GRU) which learns sequential patterns in the feature space. The resulting context of the GRU is used as model output - which has 256 dimensions - and represents the learned audio representation.

\begin{table}[t!]
\smaller
\centering
\caption{Overview of baseline audio representation. Audio-content \textit{descriptors} with corresponding acoustic \textit{categories} and feature \textit{weights} as well as the cumulative \textit{category weight} optimized for \textit{Europeana Sounds} evaluation data.}
\label{tab:features_overview}
%\begin{tabular}{p{0.19\columnwidth}|p{0.2\columnwidth}|p{0.2\columnwidth}|c|c}
\begin{tabular}{|p{0.07\textwidth}|p{0.09\textwidth}|p{0.13\textwidth}|p{0.04\textwidth}|c|}
\hline

\multicolumn{1}{|c|}{\textbf{Category}} & \multicolumn{1}{c|}{\textbf{Feature}} & \multicolumn{1}{c|}{\textbf{Description}} & \textbf{$W_{feat}$} & \textbf{$W_{cat}$} \\ \hline
                                        & \textbf{MFCC}                         & Timbre   \cite{tzanetakis2000marsyas}                                 & 23\%            &                          \\ \cline{2-4}
                                        & \textbf{SSD}                          & Spectral desc.    \cite{lidy2005evaluation}                  & 8\%             &                          \\ \cline{2-4}
\multirow{-3}{*}{\textbf{Timbre}}       & \textbf{SPEC CENT}                    & Pitch   \cite{tzanetakis2000marsyas}                                  & 8\%             & \multirow{-3}{*}{39\%}   \\ \hline
                                        & \textbf{RP}                           & Rhythm     \cite{lidy2005evaluation}                               & 18\%            &                          \\ \cline{2-4}
\multirow{-2}{*}{\textbf{Rhythm}}       & \textbf{BPM}                          & Tempo     \cite{dixon2007evaluation}                                & 7\%             & \multirow{-2}{*}{25\%}   \\ \hline
                                        & \textbf{CHROMA}                       & Harmonic Scale   \cite{tzanetakis2000marsyas}                         & 12\%            &                          \\ \cline{2-4}
\multirow{-2}{*}{\textbf{Harmony}}      & \textbf{TONNETZ}                      & Harmonic desc. \cite{harte2006detecting}                     & 12\%            & \multirow{-2}{*}{24\%}   \\ \hline
\textbf{Loudness}                       & \textbf{RMSE}                         & Loudness  \cite{tzanetakis2000marsyas}                                & 9\%             & 9\%                      \\ \hline
\textbf{Noise}                & \textbf{ZCR}                          & Noisiness     \cite{tzanetakis2000marsyas}                             & 3\%             & 3\%                      \\ \hline
\end{tabular}
\end{table}

% =================================================================
\subsection{Baseline Architecture}
\label{baseline}

The baseline approach is based on a selection of handcrafted music- and audio features. This feature-set has been specifically designed and optimized for the \textit{Europeana Sounds} audio collection \cite{Schindler2016}. Audio simlarity is calculated by late fusing the different features using different similarity metrics. The final similarity is defined the the sum of the weighted distance space. Table \ref{tab:features_overview} depicts the feature-set defined in \cite{Schindler2016}. Features are selected to describe five acoustic and musicological concepts of the heterogeneous semantics of the different collections in the \textit{Europeana Sounds} data-set. The balancing of the feature weights aims to approximate the subjective similarity.

For the baseline experiments we extract and aggregate the features listed in Table \ref{tab:features_overview} according the procedure described in \cite{Schindler2016}. To harmonize the evaluation of this paper we transformed the model to an early-fusion approach. Thus, we first standardised the value-spaces of all feature-sets separately. Then, we normalize each feature-set according their dimensionality to equalize their influence on the similarity estimation. Finally we apply weight the normalized feature-spaces according the weights of Table \ref{tab:features_overview}.

\begin{figure}[t]
  \centering \includegraphics[width=0.95\columnwidth]{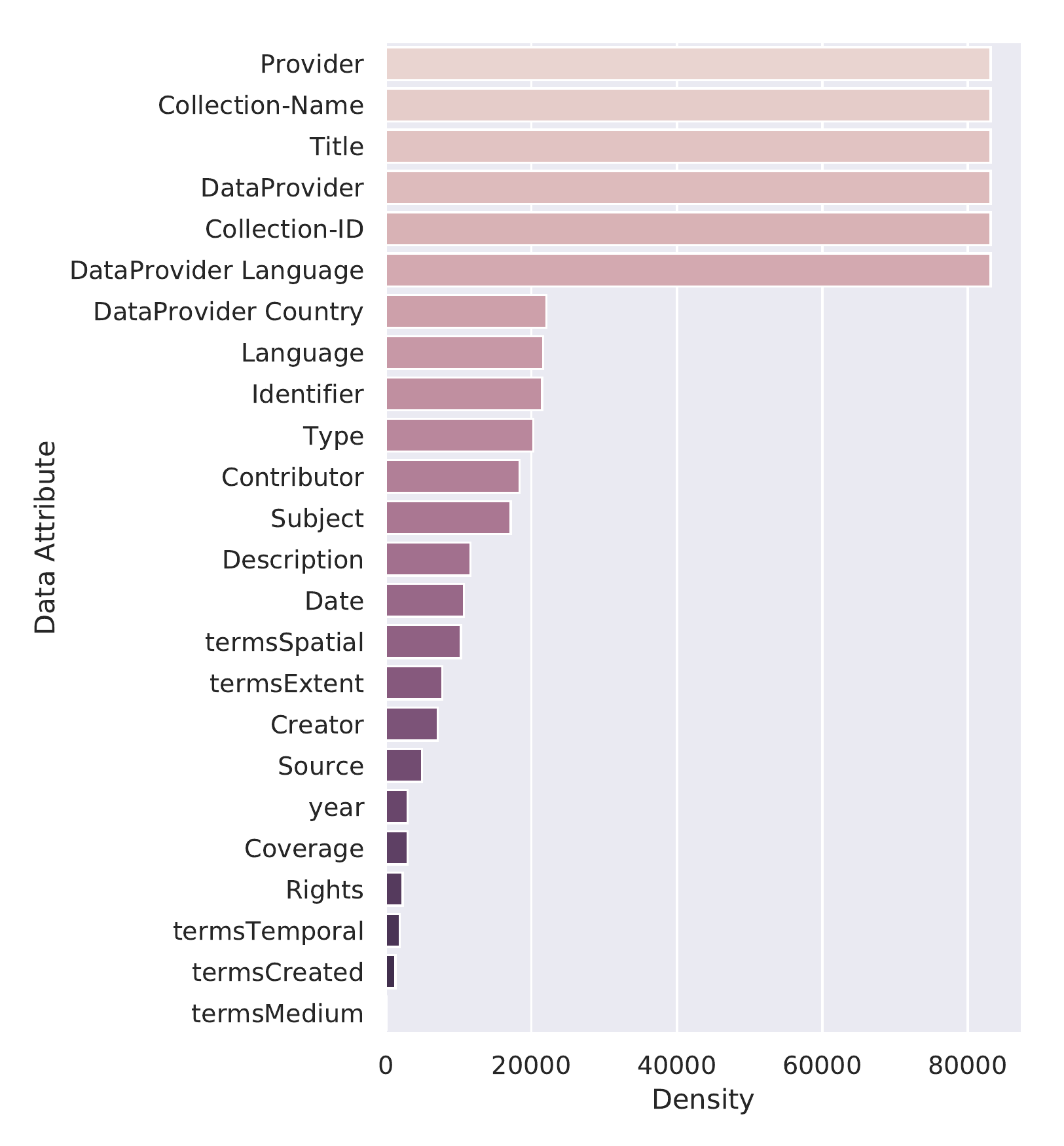}
  \caption{Metadata value density / sparsity: number provided values per metadata attribute in descending order. Top ranked attributes are densely, low ranked attributes are sparsely provided.}
  \label{fig:datadensity}
\end{figure}

\begin{figure*}[t]
  \centering \includegraphics[width=1.0\textwidth]{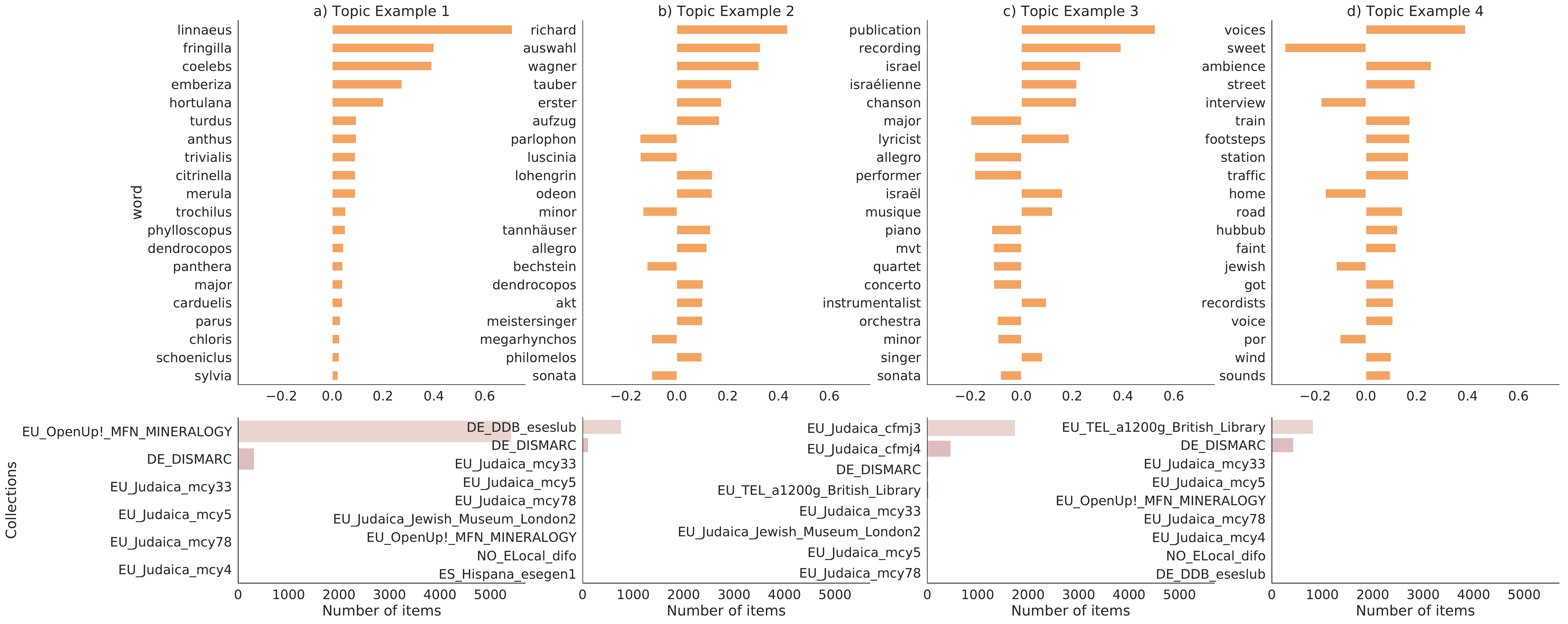}
  \caption{LSI Topic Examples visualizing the correlation between topics and collections. top - plots showing words ranked by their absolute (positive/negative) influence on the topic. bottom - number of items per collection associated with this topic. a) biological/zoological terms - mostly correlated with OpenUp! collection of animal sounds. b) terms and names of classical music and composers - mostly correlated with collection of classical music. c) terms referring to location - in this case \textit{Israel} correlated with collection of Jewish music. d) terms describing ambient sounds - mostly correlated with a collection of environmental sounds.}
  \label{fig:lsitopicexamples}
\end{figure*}

\subsection{Data} %OK

% where does the data come from? (sergiu)
The data-set used within the experimental evaluation of this paper has been developed within the scope of Europeana Sounds project\cite{schindler2016europeana}. Out of the several hundreds of thousands of audio records accessible through Europeana APIs, a subset of more than 83.000 items forms the bases of the current evaluation. The description of these audio items was collected through the Europeana Record API\footnote{\url{ https://pro.europeana.eu/resources/apis/record}}.  
Even if the representation of item description is available in a standardized knowledge graph format (using Europeana Data Model\footnote{\url{ https://pro.europeana.eu/resources/standardization-tools/edm-documentation}}), there are still several challenges to effectively use this metadata for information retrieval purposes. 

% how was it aggregated? (sergiu)
% brief description of some example collections (e.g. DISMARC, judaeica, ..) (sergiu)
The Europeana Records are collected from various Institutions from all over the Europe and pre-processed by so called national or thematic aggregators.
As shown in Figure \ref{fig:dataoverview}, the records included within the evaluation dataset were originally provided by 30 different institutions (also called data providers) and included in 26 different data collections. % (where the collection names typically include also the name of the data aggregator). 
The \textit{Europeana Sounds} project, with \textit{Dismarc} music collection and several other non music collections, is the main thematic aggregator for sound content. It is followed by the Jewish Heritage Network, the aggregator of Judaica collections, which include many traditional songs collected from Jewish communities that were leaving in different European countries. However, many records were submitted to Europeana by national aggregators and previous research projects.  

% what are the challenges, pros and cons? (sergiu)
The large variety and data sparsity within this evaluation corpus represent a great challenge for the efficiency of the proposed approach. This variation is encountered with respect to the type of content, the used categorization schemes and with respect to the distribution of audio features. The largest part of sound content is represented by music records, however an important part of the data set includes radio news, public speeches or language dialects recordings, environmental (e.g. city noise) or biodiversity sounds (e.g. bird twittering), etc. In the case of music records, there is a high variety of music genres, from traditional to classical music, from love songs to rock, from instrumental to single voice singer, etc. As consequence, the evaluation dataset builds a sparse data matrix, except for the fields that are enforced as mandatory through the Europeana Data Model (see Figure \ref{fig:datadensity}).  

Another challenge of the evaluation corpus is represented by the multilingualism of the metadata, more than 40 European languages being now used to describe the Europeana records. Even if the several data fields contain very precise keywords describing the audio content, they have limited usage for similarity search due to their language distribution. 
%In order to overcome this drawback, we used Google Translation to collect all relevant information in English language. This facilitates the clustering of similar audio items by their descriptive terms using an automatic topic detection approach. 
In Figure \ref{fig:lsitopicexamples} we showcase the composition of the topics and indicate the influence of individual keywords. Through the topic composition, a human user can easily recognize topics relating to biodiversity and animal sounds, classical music and composers, locations, ambient sounds, etc.   
Europeana data collections are meant to group records that share certain properties in common, the enclosed records being similar to each other in a broader sense. However, in many cases, data collections group items from different institutions and cover several topics. The correlation between topics and data collections is indicated in Figure \ref{fig:lsitopicexamples}.              

% [ 'dcContributor', 'dcCoverage', 'dcCreator', 'dcDate', 'dcDescription', 'dcIdentifier', 'dcSource', 'dcSubject', 'dcTitle', 'dcType', 'dctermsCreated', 'dctermsMedium', 'dctermsSpatial', 'dctermsTemporal', 'dctermsExtent', 'year', 'europeanaCollectionName', 'edmCountry', 'edmLanguage', 'dcLanguage', 'edmDataProvider', 'edmProvider']

From the full description of audio items, 22 metadata elements were taken for computing item similarities. Some of these elements indicate the provenance and the aggregation process of audio content. The Institution that owns the content is named within the \textit{Data Provider} element, while the \textit{Provider} indicates the organization aggregating the content in the collection defined though the \textit{Collection-Name} and \textit{Collection-id}. The country and the language used by the contributing institution are also available in the metadata (i.e. \textit{DataProvider Language}, \textit{DataProvider Country}. The elements describing the audio content include the a mandatory \textit{Title} and optional \textit{Subject} categorizations, \textit{Type} of content and textual \textit{Description}. 
%The authors and persons contributing to the creation of the audio recordings are indicated within the \textit{Creator} and \textit{Contributor} elements. 
Quite often, within the contributor fields the role of the person is also indicated, or in case of orchestras, the music instrument might be available as well.
When known, the \textit{Date} and the \textit{Year} are available directly in the medata elements. The approximate date and location can be indicated either through the \textit{Coverage} or through the specialized \textit{termsTemporal} and \textit{termsSpatial}, \textit{termsCreated} elements. For some items the the storage medium (\textit{termsMedium}) and the original work from which the current object was derived are also indicated. All these data elements contain information that has correlation with the characteristics of the audio content.    

\subsection{Data Pre-processing} %OK
\label{dataprep}

\subsubsection{Text-data Pre-processing} %OK

This section refers to the text-input to build the LSI models. All meta-data attributes are concatenated to a single string per instance using white-spaces as separators. The entire text content is converted to lower-case and HTML tags, sequences and hyperlinks are removed. Year dates are mapped to decades and centuries and translated to Roman numerals. Further, numbers and punctuation's are removed. After tokenization, stopwords are removed for the languages \textit{English, German, Italian, French} and \textit{Romanian}. Finally, words $w$ with $|w| <= 2$ are also removed. 

\subsubsection{Audio Data Pre-Processing} %OK

All audio-files of the collection are re-sampled to 44.100 Hz and single-channel converted. An audio segment $s$ of length $|s| = 6$ seconds is read from an audio-file $a$ using an offset of $o = 5$ seconds to avoid silent sections at the beginning of a recording as well as fade-in effects. If $|a| <= |s| + o$ the offset is reduced to $o = |a| - |s|$. If $|a| < |s|$, the audio-file is shorter than the expected sequence length and the missing content is zero-padded.
Short-time Fourier Transform (STFT) with a Hanning-windowing function and a 4096 samples window size with 50\% overlap is applied. The resulting Spectrogram is transformed to the log-scaled Mel-space using 80 Mel-filters and cut-off frequencies of 16Hz (min) and 18.000Hz (max). The final shape of the DNN input matrix shape is 80x130x1. Instead of normalizing the feature-space, we add a batch-normalization layer on top of the network (see Figure \ref{fig:modelarchitecture}).

\begin{figure}[t]
  \centering \includegraphics[width=1.0\columnwidth]{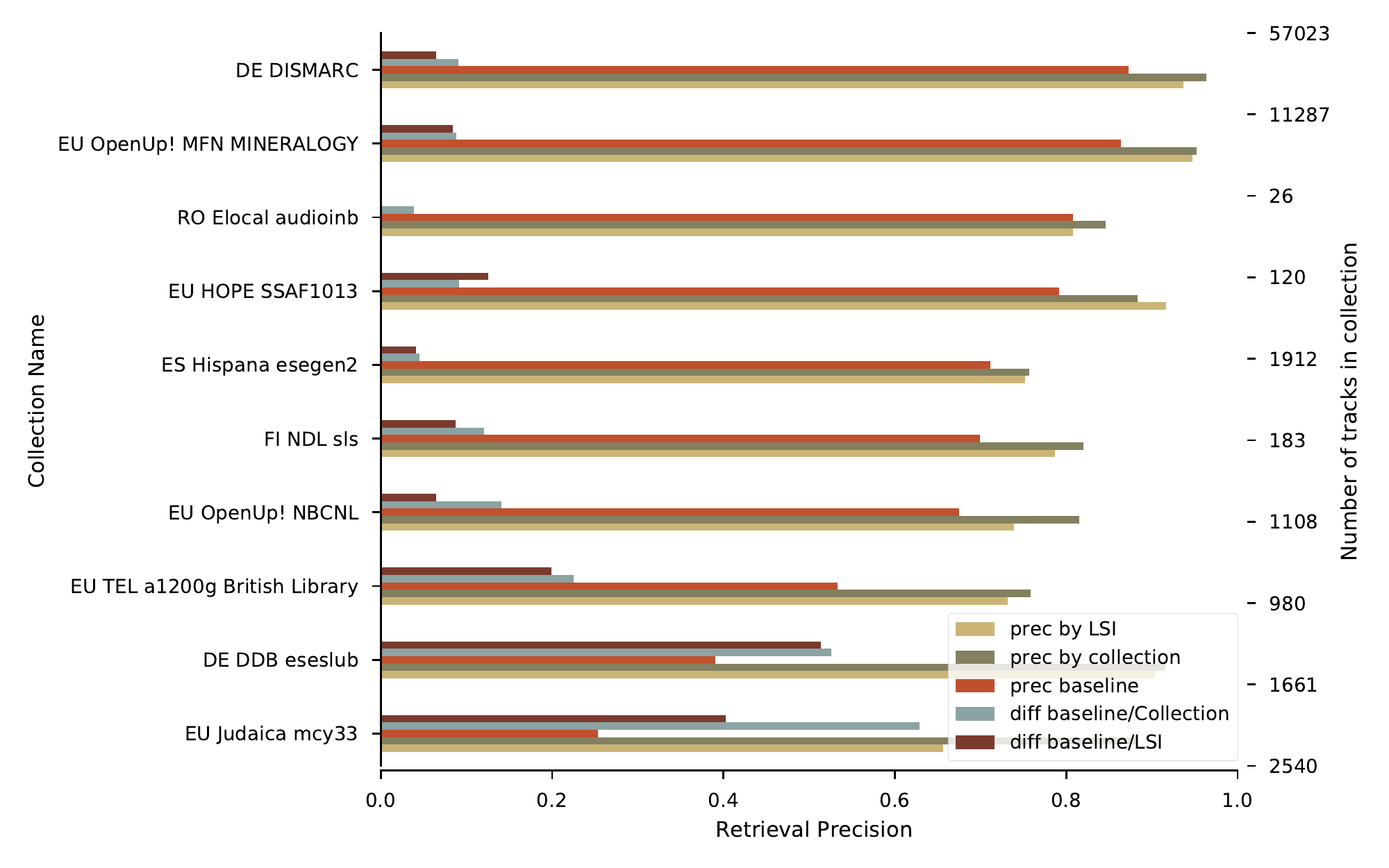}
  \caption{Evaluation results \textit{Similarity by collection membership} measured in retrieval precision at cut-off 1 and number of tracks of the corresponding collection (right Y-axis). Global mean precision at different cut-offs presented in Table \ref{tab:results_avg_prec_collection}.}
  \label{fig:results_collections}
\end{figure}

% =================================================================
\section{Results}
\label{sec:results}
% =================================================================

Results are calculated identically for all experiments. First, the handcrafted features or trained models are applied to the audio collection to embed it into a feature-space. On this, nearest-neighbor search is performed using Euclidean distance as similarity function. To compare the results and evaluate the approaches, retrieval precision at different cut-offs (1,2,3,5,10,50,100) is assessed. The model performance is then evaluated according two criteria. 

\begin{enumerate}
    \item \textbf{Collection Similarity:} collections provided by partner libraries to the Europeana have been carefully aggregated, are coherent and share common attributes and characteristics. 
    \item \textbf{Term Similarity:} content related terms such as music instruments, composers, music genres or animal species are matched in the metadata. These terms span across several collections and calculate their entropy to assess if they are evenly distributed or skewed towards a certain collection.
\end{enumerate}

%Results are presented in Figure \ref{fig:results_collections} and \ref{fig:results_terms} and are discussed as follows.

% -----------------------------------------------------------------
\paragraph{\textbf{Track-relatedness by Collection Membership:}}

\begin{figure}[t]
  \centering \includegraphics[width=1.0\columnwidth]{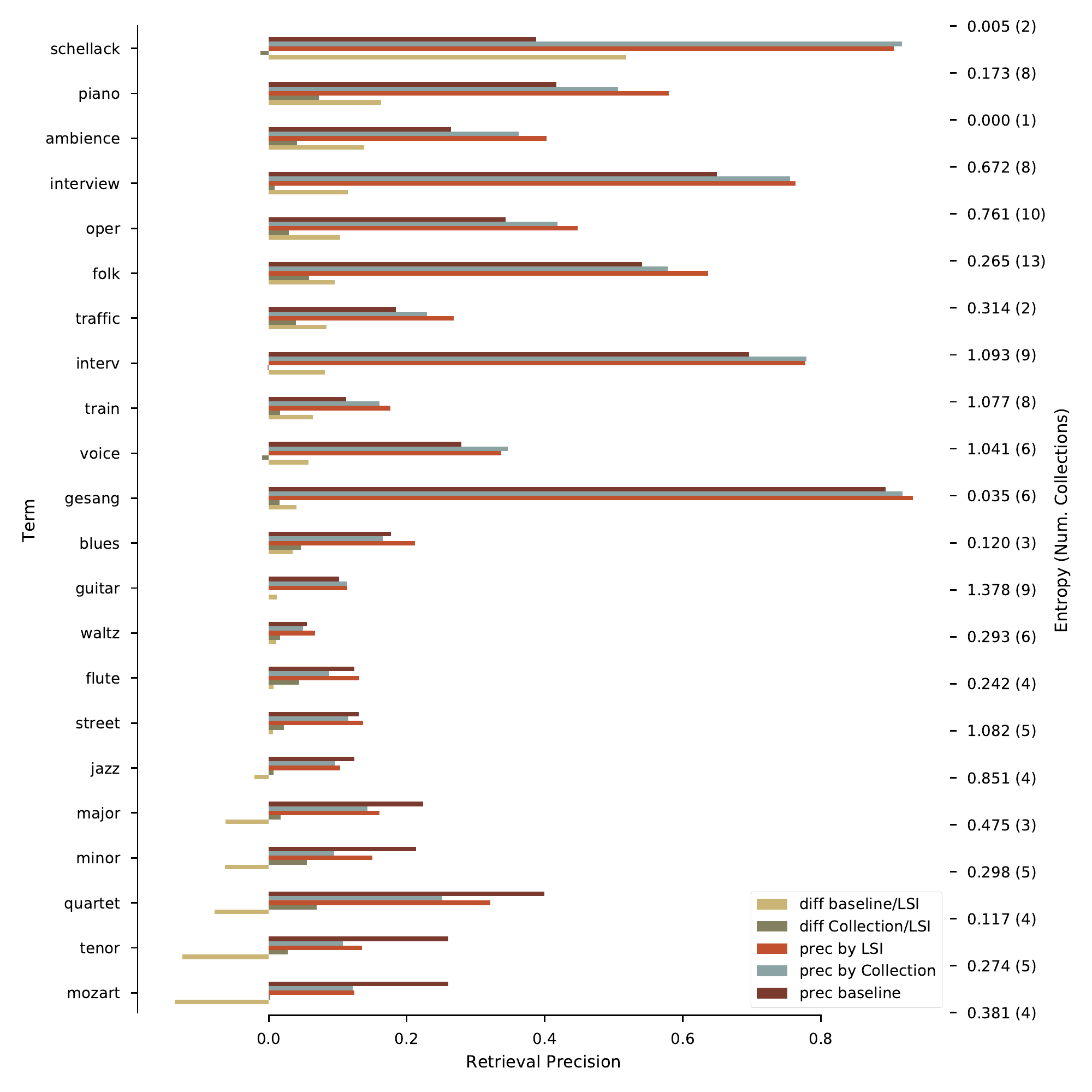}
  \caption{Evaluation results \textit{Similarity by content related terms} measured in retrieval precision at cut-off 1. Right Y-axis shows term entropy across the number of collections. Global average precision at different cut-offs presented in Table \ref{tab:results_avg_prec_collection}.}
  \label{fig:results_terms}
\end{figure}

The first evaluation is based on common approaches to representation learning which use membership to a class \cite{hoffer2015deep}, label \cite{schindler_knees:cbmi:2019} or identity \cite{schroff2015facenet,park2018representation} to select positive and negative examples for triplet based neural networks.
As can be observed in Figure \ref{fig:results_collections} this approach tends to learn representation which focus on collection related features and acoustic artifacts. From this perspective, this approach seems to outperform the baseline as well as the proposed approach, but, although, the collections of the dataset have been well aggregated by professional librarians, the learned representation do not generalize towards to a global acoustic similarity, as can be observed in Figure \ref{fig:results_terms}. The term-based precision values for the model trained on track-relatedness by collection membership differ recognizably from the LSI based approach, especially for evaluation terms with high entropy values.

\paragraph{\textbf{Track-relatedness by LSI-Topic Similarity:}}

The potential of the proposed LSI-based representation learning approach is shown by the decrease of precision in the collection-based and the increase in the term-based evaluation. This indicates that the learned representation capture  characteristics which facilitate acoustic similarity estimations globally, across the collections of the dataset.
For some terms the representations learned from free text outperform the baseline significantly, such as for the acoustic characteristics of digitized shellac recordings which are difficult to model with handcrafted audio features. It can generally be observed that the representation learned by our approach improves over the baseline in terms of capturing general acoustic characteristics such as ambient sounds, sound producing objects such as instruments or machines as well as human voice. Regarding the first baseline following the feature-based approach presented in \cite{Schindler2016}, our proposed approach does not capture music related characteristics accordingly. An explanation for this could be, because only 6 dimensions of the feature-set of the baseline approach are general audio features wheres the remaining 1671 dimensions belong to music features. Regarding the second baseline using collection membership as track-relatedness measure, the LSI-based track-relatedness approach improved over almost all term-based results.

\begin{table*}[t]
%\smaller
\centering
\caption{Overview of baseline audio representation. Weighted composition of state-of-the-art audio and music content \textit{Features}. The audio-content \textit{descriptors}, their corresponding acoustic \textit{categories}, their assigned feature \textit{weight} as well as the cumulative \textit{category weight}.}
\label{tab:results_avg_prec_collection}
\begin{tabular}{l|r|rrrrrrr}
\toprule
{} & Dim &     1 &     2 &     3 &     5 &    10 &    50 &   100 \\
\midrule
baseline      & 1677 & \textbf{0.288} & \textbf{0.253} & \textbf{0.235} & 0.213 & 0.188 & 0.139 & 0.121 \\
by Collection &  256 & 0.232 & 0.216 & 0.208 & 0.196 & 0.182 & 0.156 & 0.143 \\
by LSI        &  256 & 0.264 & 0.241 & 0.231 & \textbf{0.218} & \textbf{0.201} & \textbf{0.170} & \textbf{0.158} \\
\bottomrule
\end{tabular}
\end{table*}

\paragraph{\textbf{Discussion:}}

Audio similarity is generally difficult to evaluate. Usually, categorical ground truth assignments are used alternatively and precision is defined on retrieving tracks belonging to the same category \cite{knees2016music}. In this paper we describe an approach to learn an audio representation for similarity estimation from multi-lingual free-text under the absence of ground-truth data. Thus, we approach the evaluation from two perspectives: the categorical view using collection memberships and terms in the metadata describing acoustic properties. With this approach we are able to show that the LSI based approach to audio representation learning provides similar retrieval precision to the feature-based baseline approach up to a cut-off of 3 (see Table \ref{tab:results_avg_prec_collection}). From a cut-off of 5 our learned representations with 256 dimensions exceed the baseline with 1677 dimensions. Thus, the size of the feature-space is reduced by a factor of 6.5 at consistent performance.

% TODO & TODO & TODO & TODO & TODO & TODO & TODO & TODO & TODO & TODO & TODO & TODO & TODO & TODO & TODO
% TODO finsh table 
% TODO & TODO & TODO & TODO & TODO & TODO & TODO & TODO & TODO & TODO & TODO & TODO & TODO & TODO & TODO
%\begin{table}[t!]
%%\smaller
%\centering
%\caption{Overview of collection used for evaluation 1. Rows contain collection name, language(s), short description and %number of tracks.}
%\label{tab:overview_collections}
%
%\begin{tabular}{lr}
%\toprule
%\textbf{Collection description} &  Num. Tracks \\
%\midrule
%\textbf{DISMARC} (DE): \cite{koch2009} large music collection, &  56945 \\
%various genres & \\
%\textbf{OpenUp! MFN MINERALOGY} (EU):     &  10453 \\
%\textbf{Elocal audioinb} (RO):            &  26 \\
%\textbf{HOPE SSAF1013} (EU):              &  120 \\
%\textbf{Hispana esegen2} (ES):            &  1912 \\
%\textbf{NDL sls} (FI):                    &  183 \\
%\textbf{OpenUp! NBCNL} (EU):              &  1097 \\
%\textbf{TEL a1200g British Library} (EU): &  976 \\
%\textbf{DDB eseslub} (DE):                &  1661 \\
%\textbf{Judaica mcy33} (EU): traditional jewish music,\\ old recordings  &  2540 \\
%\bottomrule
%\end{tabular}
%
%\end{table}

\section{Conclusions and Future Work}
\label{sec:conclusions}

We introduced a novel approach to unsupervised audio representation learning. We showed how to estimate track-relatedness from unstructured multilingual free-text by projecting the semantic data into a vector space using Latent Semantic Indexing. This track-relatedness is used for online triplet selection to train a triplet deep neural network which cross-learns an audio representation from the text modality. We showed that the representations learned perform similar to the baseline, respectively exceeding the baseline in similarity retrieval precision at higher cut-offs at only 15\% of the baseline's feature vector length.

\medskip
\small
\textbf{Acknowledgements} \textit{This article has been made possible partly by received funding from the European Unions Horizon 2020 research and innovation program in the context
of the VICTORIA project under grant agreement no. SEC-740754}

\bibliographystyle{ACM-Reference-Format}
\bibliography{references}

\end{document}